\documentclass[10pt]{article}
\usepackage{amsfonts}
\usepackage{amsthm}

\usepackage{xspace,colortbl}
\usepackage[latin1]{inputenc}
\usepackage[T1]{fontenc}
\usepackage{graphicx}
\usepackage{graphics}
\usepackage{verbatim}
\usepackage{bookman}
\usepackage{amsmath}
\usepackage{amssymb,latexsym}
\usepackage{amscd}
\usepackage{mathrsfs}
\usepackage{pifont}
\usepackage{color}

\usepackage{graphicx}

\newcommand{\sss}{\setcounter{equation}{0}}
\newtheorem{theorem}{THEOREM}[section]

\newtheorem{remark}[theorem]{REMARK}

\newtheorem{prop}[theorem]{PROPOSITION}

\newtheorem{assumption}[theorem]{ASSUMPTION}

\newcommand{\ere}{ {\mathbb R}}

\newcommand{\ese}{{\mathbb S}}

\def\beq{\begin{equation}}
\def\ene{\end{equation}}
\def \ds {\displaystyle}
\newcommand{\bull}{\hfill $\Box$}
\def\x{\mathbf x}
\def\y{\mathbf y}
\def\p{\mathbf p}
\def\q{\mathbf q}

\begin{document}
\baselineskip=20 pt
\parskip 6 pt

\title{Universality of Entanglement Creation in Low-Energy Two-Dimensional Scattering
\thanks{ PACS Classification (2010): 03.67.Bg, 03.65.Nk, 03.65.Db.  Mathematics Subject Classification(2010):81P40, 81U05, 35P25.}
\thanks{ Research partially supported by
 CONACYT under Project CB-2008-01-99100.}}

 \author{ Ricardo Weder\thanks {Fellow, Sistema Nacional de Investigadores.}
\\
Departamento de F\'{\i}sica Matem\'atica.\\ Instituto de Investigaciones en Matem\'aticas Aplicadas  y en
 Sistemas. \\Universidad Nacional Aut\'onoma de M\'exico. Apartado Postal 20-726,
M\'exico DF 01000.\\
Electronic mail: weder@unam.mx  }
\date{}
\maketitle

\vspace{.5cm}
 \centerline{{\bf Abstract}}
\noindent
We prove that the entanglement created in the low-energy scattering of two particles  in two dimensions is given by a universal coefficient that is independent of the interaction potential. This is strikingly different from the three dimensional case, where it is proportional to the total scattering  cross section. Before the collision the state  is a product of two normalized Gaussians. We take the purity as the measure of  the entanglement after the scattering. We give a rigorous  computation, with error bound, of the leading  order of the  purity at  low-energy. For a large class of potentials, that are not-necessarily  spherically symmetric, we prove that the low-energy behavior of the purity,  $\mathcal P$, is universal.
 It is given by  $\mathcal P= 1-\, \frac{1}{(\ln (\sigma/\hbar))^2} \, \mathcal E$, where $\sigma$ is the variance of the Gaussians and the entanglement coefficient, $\mathcal E$, depends only on the masses of the particles and not on the interaction potential.  There is a strong dependence  of the entanglement  in the difference of the masses. The minimum is when the masses are equal, and  it increases strongly with the difference of the masses.

\vspace{.5cm}

\bigskip
\noindent
\section{Introduction}\sss
  In this paper we consider the creation of entanglement in the low-energy scattering of two  particles without spin in two dimensions.  The  interaction between the particles
  is given by a general potential that is not required to be spherically symmetric. Before the scattering the particles are in an incoming
  asymptotic state that is a product of two Gaussians. After the scattering the particles are in an outgoing asymptotic state that  is not a product state.  The problem that we solve  is to compute the loss of purity, due to the entanglement with the other, that is produced by the collision.

 In the configuration representation the Hilbert space of states for the two particles is  $\mathcal H:= L^2(\ere^4)$. The  dynamics of the
 particles is given by the Schr\"odinger equation, \beq\label{1.1} i \hbar
\frac{\partial }{\partial t}\varphi(\x_1,\x_2)= H\varphi(\x_1,\x_2).
 \ene
 The Hamiltonian is the following operator,
 \beq \label{1.2}
  H=H_0+V(\x_1-\x_2),
  \ene
where $H_0$ is the free Hamiltonian,
\beq\label{1.2b}
 H_0:=
-\frac{\hbar^2}{2m_1} \Delta_1 -\frac{\hbar^2}{2m_2} \Delta_2.
 \ene
The operator  $\Delta_j$, is the Laplacian in the coordinates $\x_j,
j=1,2$, of particle one and two. By  $\hbar$  it is denoted Planck's
constant. Furthermore, $m_j,  j=1,2$, are, respectively, the mass of
particle one and two, The   interaction potential is  a real-valued
function, $V(\x)$, defined for $ \x\in\ere^2$. We suppose that the
interaction depends on the difference of the coordinates
  $\x_1-\x_2$, but we do not require the spherical symmetry of the potential.
  We consider a general class of potentials that satisfy mild assumptions on its  decay at infinity and on its
  regularity:
  \begin{assumption}\label{ass1.1}
\beq \label{1.3}
(1+|\x|)^{\beta} V(\x)  \in L^2(\ere^2),   \quad \hbox{\rm for some}\, \beta >11.
\ene
\end{assumption}
Under this assumption $H$ is a self-adjoint operator.

 We also suppose that at zero energy there is neither an eigenvalue nor a resonance (half-bound state), for the Hamiltonian of
  the relative motion $H_{\mathrm rel}:=- \frac{\hbar^2}{2m}\,\Delta_{\x} +V(\x)$ where $\x \in \ere^2$ is the relative distance  and $m$ is the reduced mass
   $m:= m_1\,  m_2/(m_1+m_2)$.
  A zero-energy resonance (half-bound state) is a bounded solution to $H_{\mathrm rel}\varphi=0$  that is not in $L^2(\ere^2)$. See \cite{jn}
  for a precise definition.  For generic potentials $V$ there  is neither a resonance nor an eigenvalue at zero for $H_{\mathrm rel}$. That
   is to say, if we consider the potential $\lambda V$ with a coupling constant $\lambda$, zero can be a resonance and/or an eigenvalue for
   at most a finite or denumerable set of $\lambda$'s without any finite accumulation point.

We study our problem  in the  center-of-mass frame. We consider an incoming asymptotic state that is a product of two normalized Gaussians,
given in the momentum representation by,
\beq\label{1.5}
\varphi_{\rm in,\p_0}(\p_1,\p_2):=  \varphi_{\p_0}(\p_1) \,  \varphi_{-\p_0}(\p_2),
\ene
with,
\beq\label{1.6}
\varphi_{\p_0}(\p_1):=  \frac{1}{(\sigma^2\pi)^{1/2}}   e^{- (\p_1-\p_0)^2 /  2\sigma^2},
\ene
where $\p_i,i=1,2$ are, respectively, the momentum of particles one and two.

In our incoming asymptotic  state (\ref{1.5}) particle one has mean momentum $ \p_0$ and particle two has  mean momentum $ -\p_0$.
 Both particles have the same variance, $\sigma$, of the momentum distribution. As we suppose that the scattering takes place at the origin at
 time zero, the   average position of both particles  is zero in the incoming asymptotic state   (\ref{1.5}).
The outgoing  asymptotic state of the two particles, $\varphi_{\rm out, \p_0}$, after the scattering process, is   given by,

\beq\label{1.7} \varphi_{\rm out,\p_0}(\p_1,\p_2): = \left( \mathcal
S(\p^2/2m)  \varphi_{\mathrm in, \p_0 }\right)(\p_1,\p_2). \ene
Here,  $\p:= \frac{m_2}{m_1+m_2}\p_1-\frac{m_1}{m_1+m_2}\p_2$ is the
relative momentum, and $ \mathcal S( \p^2/2m )$ is the scattering
matrix for the relative motion.

The  measure of entanglement of a pure-bipartite state  that we use is the purity. Namely,  the trace of the square of
the reduced density matrix of one of the particles, that is obtained by taking the trace on the other particle of  the density matrix of the
pure state. Note that the purity of a product state is one.

In the outgoing asymptotic state  $\varphi_{\rm out, \p_0}$ the
purity is given by,
 \beq\label{1.8} {\mathcal P}(\varphi_{\rm out,
\p_0})= \int \, d\p_1 \,d\p_1' \,d\p_2 \,d\p_2'
\,\varphi_{\rm out, \p_0}(\p_1,\p_2)
 \,\overline{\varphi_{\rm out, \p_0}(\p_1', \p_2)} \,\varphi_{\rm out, \p_0}(\p_1',\p_2') \,\overline{\varphi_{\rm out, \p_0}(\p_1,\p_2')}.
\ene
 Remark that  $\varphi_{\rm out, \p_0}$ is not a product state, and that its purity is smaller than
 one, because the relative momentum, $\p$, depends on $\p_1$ and on
$\p_2$. This implies that the collision has created entanglement
between the two particles.

To be in  the low-energy regime the following two conditions have to
be satisfied. 1. The mean  relative momentum $\p_0$ has to be small.
2. The variance $\sigma$ has to be small. Note that  even if the
mean relative momentum $\p_0$ is small, if $\sigma$ is large the
incoming asymptotic state   $\varphi_{\rm in, \p_0}$  has a big
probability of having large momentum.

Let us designate  by $\varphi_{\rm in}$  the incoming asymptotic state with mean relative momentum $\p_0=0$. The corresponding outgoing asymptotic
 state  is  $\varphi_{\rm out}:=   \mathcal S(\p^2/2m)  \varphi_{\mathrm in} $.

Let us designate by
 \beq\label{1.8b}
 \mu_i:= \frac{m_i}{m_1+m_2}, i=1,2,
\ene
the fraction of the mass of the $i$ particle to the total mass.

In Section 3 we  rigorously prove the following results on the
leading order of the purity  at low energy.

\beq\label{1.9}
 \mathcal P(\varphi_{\rm out, \p_0})= \mathcal P(\varphi_{\rm out})+\frac{|\p_0|}{\sigma}\,\, O\left(\frac{1}{|\ln(\sigma/\hbar+|\p_0|/\hbar)|^{2}}\right), \qquad \rm{as} \, \sigma/\hbar+|\p_0|/\hbar \rightarrow 0,
 \ene
 \beq\label{1.9b}
 \mathcal P(\varphi_{\rm out})=1-  \frac{1}{(\ln(\sigma/\hbar))^2} \mathcal E(\mu_1) + O\left(\frac{1}{|\ln(\sigma/\hbar)|^3}\right),  \qquad   \rm{as}\, \sigma/\hbar \rightarrow 0,
 \ene
where $\mathcal E(\mu_1$) is the entanglement coefficient,
\beq\label{1.9c}
\mathcal E(\mu_1):= \ds
\frac{2 \pi^2}{1+(2\mu_1-1)^2}\,\, \left[1+ \sqrt{1+(2\mu_1-1)^2}\right]- \frac{2}{\pi} [J(\mu_1,1-\mu_1)+ J(1-\mu_1,\mu_1)],
\ene
with
\beq \label{1.9d}\begin{array}{l}
J(\mu_1,\mu_2):=  \int d\q_2\left[ \,\,\int d\q_1\, \hbox{ Exp}[-\frac{1}{2}(\mu_1^2+\mu_2^2)(\q_1+\q_2)^2-(\mu_2\q_1-\mu_1\q_2)^2 - \q_1^2/2] \,\,\cdot \right.\\\\ \left.
\ds I_0(|\mu_1-\mu_2|\, |\q_1+\q_2|\,|\mu_2\q_1-\mu_1\q_2|)\right]^2,
 \end{array}
\ene
where $I_0$ is the modified  Bessel function \cite{gr}.

We have that $\mathcal E(\mu_1)= \mathcal E(1-\mu_1)$. This is a
consequence of the invariance of $\mathcal P(\varphi_{\rm out})$
under the exchange of particles one and two.

 Note that $J(1/2,1/2)= \pi^3$. In the appendix we prove explicitly that  $J(1,0)= 16.6377$. For  $\mu_1 \in [0,1]\setminus\{1/2,1\}$ we
 compute  $J(\mu_1,1-\mu_1)$ numerically using Gaussian quadratures.

 The entanglement coefficient $\mathcal E(\mu_1)$ is  universal in the sense  that is independent of the interaction potential. Of course, this is only true if the potential is not identically zero, see  the low-energy estimate for the scattering matrix given in (\ref{2.8c}). It follows from  \eqref{1.9b} that the creation of entanglement is a second-order effect.
 Note that in the second-order term in  (\ref{2.8c}) it  appears the scattering length, $a$, that depends on the potential. However, the contributions to the leading order of the purity that depend on the scattering length cancel each other.
 
The universality of the entanglement at low-energy in two dimensions is strikingly different from the three dimensional case that we previously studied in \cite{we},  where the entanglement created by the  collision  is proportional to the total scattering  cross section. It is a natural question to ask what are  the physical reasons why the results are so different in two and in three dimensions. This is certainly a non trivial issue. I propose the following answer. First note that constraining  particles to live in two dimensions is a strong requirement. It dramatically changes the kinematics for low energy. This can be seen, for example, in the well known logarithmic divergence of the free Green's function, that is absent in three dimensions. This difference in the kinematics also affects the dynamics of the particles, specially at low energy, where the asymptotics of the scattering matrix  is fundamentally different in two and in three dimensions. Both in two and in three dimensions the creation of entanglement is due to kinematical effects that depend on the masses of the particles and to dynamical effects that depend on the potential of interaction. In three dimensions these effects are of the same order at low energy. However, when the particles are constrained to two dimensions, also the dimension of the phase space is reduced,  and the  kinematical  effects play a dominant role at low energy, and, in consequence,  the details of the potential do not play a role in the leading order of the entanglement. In intuitive physical terms this is the main reason for the universality of the low-energy entanglement in two dimensions.

 Table 1  and Figure 1 show that -as in three dimensions \cite{we}- the entanglement  coefficient depends strongly in the difference of the
 masses. The minimum is taken for $\mu_1=0.5$, i.e. for equal the masses, and it strongly increases  with the difference of the masses,
 when $\mu_1$ tends to one. For a physical interpretation of this
 fact see \cite{we}.

Universal results, like the one of this paper,  are certainly of independent interest. They point out to deep fundamental physical issues  and they do not need to be justified by applications. However, our result has important potential applications. For example, in the creation of entanglement  in two-dimensional systems where scattering is essential, like ultracold  particles, or solid state devices. In this situation, our result shows that it is possible to produce entanglement in experiments where the particles interact very weakly, provided that the difference in the masses is large. Furthermore, we provide a formula for the created entanglement that can be verified experimentally. This, of course, requires experiments  where  it is possible to experimentally  measure the  entanglement, what is an  issue on itself.           

 There are many other reasons to study the entanglement creation in scattering
 processes. Entanglement is a central issue in quantum information
 and scattering is fundamental in all areas of physics. For a
 detailed  physical motivation and for other possible  applications see
 \cite{we}.

For previous results in the generation of entanglement in scattering processes  in one dimension, mainly for potentials with explicit solution,  see \cite{sj}, \cite{hs}, and the references quoted in these papers.  Actually these papers do not obtain low-energy estimates of the creation of entanglement in one dimension that can be compared to our results  in the three dimensional case in  \cite{we} or to our two-dimensional results in this paper. In fact, a precise analysis of the low-energy entanglement creation in one dimension is an open problem that we intent to study in future investigations. Furthermore,
  \cite{da},  \cite{dfa},  \cite{afft}, and  the references quoted in these papers, study  a system consisting of heavy and light particles. They study the asymptotic dynamics and  the decoherence that is produced on the heavy particles  by the collision with the light particles in the limit when  the  mass ratio is small. Note that this is a  different problem from the one that we discuss here and in \cite{we}.  Furthermore, the loss of quantum coherence that is induced on  heavy particles by the interaction with light ones  has attracted  a great deal of attention. For example, see  \cite{jz}, and \cite{gf}.

The paper is organized as follows. In Section 2 we  study the low-energy asymptotics of  the scattering matrix for the relative motion of the particles. In Section 3 we give the proof of the  results in the creation of entanglement. In Section 4 we give our conclusions. In the Appendix we explicitly evaluate integrals that we need in Section 3. Along the paper we denote by $C$ a generic positive constant  that does not necessarily have the same value in different appearances.

\section {Scattering at Low-Energy in Two Dimensions}
\sss

We denote by $\hat{\mathcal H}:= L^2(\ere^4)$ the state space in the momentum representation.  The momentum of the particles  one and two are, respectively, $\p_1,\p_2$.
It  is convenient to take as coordinates in the momentum representation the momentum of the center of mass and the relative momentum,
\beq
\begin{array}{l}\label{2.1}
 \p_{\rm cm}:= \p_1+\p_2, \\ \\
\p:= \frac{m_2 \p_1- m_1\p_2}{m_1+m_2}.
\end{array}
\ene
The state space in the momentum representation  factorizes as a tensor product,
\beq\label{2.2}
\hat{\mathcal H}= \hat{\mathcal H}_{\rm cm} \otimes \hat{\mathcal H}_{\rm rel},
\ene
where $\hat{\mathcal H}_{\rm cm}=L^2(\ere^2), \hat{\mathcal H}_{\rm rel}:=L^2(\ere^2)$ are, respectively, the state spaces in the momentum representation for the center-of-mass motion and the relative motion.

Since the potential depends on the difference of the coordinates of particles one and two the scattering matrix for the system decomposes as the tensor product  $I_{\mathrm cm} \otimes  {\mathcal S}(\p^2/2m)$ of the identity on
$\hat{\mathcal H}_{\rm cm}$ times  the scattering matrix for the relative motion,  ${\mathcal S}(\p^2/2m)$,  in  $\hat{\mathcal H}_{\rm rel}$, where $m$ is the relative mass,
\beq\label{2.3}
m:= \frac{m_1 \, m_2}{m_1+m_2}.
\ene

The scattering matrix ${\mathcal S}(\p^2/2m)$  is a unitary operator in $L^2(\ese^1)$ for each  $ \p^2/2m\in (0,\infty)$,
where we denote by $\ese^1$ the unit circle in  $\ere^2$.

We introduce some notation that we need. We denote $v:=  \sqrt{|V(x)|}$ . Let $P,Q$ be the projector operators in $L^2(\ere^2)$,

\beq\label{2.4}
P:= \frac{1}{\alpha}v(\x) \left(\cdot, v\right), \quad Q:= 1-P,
\ene
where,
\beq \label{2.5}
\alpha:= \int_{\ere^2} |V(\x)|\, d\x.
\ene
Furthermore,
\beq \label{2.6}
U(\x):=\,\left\{ \begin{array}{lr} 1, \quad\hbox{\rm if} \,\,V(\x)\geq 0,\\ -1 \,\,\quad \hbox{\rm if} \,\,V(\x) < 0.
\end{array} \right.
\ene
By $\bf M_{00}$ we denote the integral operator with kernel,
\beq \label{2.7}
M_{00}(\x,\y):= U(\x)\, \delta (\x-\y)-\frac{1}{2 \pi} \frac{2m}{ \hbar^2}\, v(\x)\, \ln\left( \frac{e^{\gamma}\,|\x-\y|}{2} \right)\, v(\y),
\ene
where $\gamma$ is Euler's constant. Moreover, by ${\bf N_{00}}$ we  denote the integral operator with kernel,
\beq\label{2.8}
N_{0,0}(\x, \y):= U(\x)\, \delta (\x-\y)-\frac{1}{2 \pi} \frac{2m}{ \hbar^2}\, v(\x)\, \ln\left( |\x-\y| \right)\, v(\y),
\ene
and
\beq \label{2.8a}
D_0:= (Q {\bf M_{00}}Q)^{-1}, \, \hbox{\rm   a bounded operator,}\,\,  Q L^2(\ere^2) \rightarrow Q L^2(\ere^2).
\ene
The assumption that $0$ is neither a resonance nor an eigenvalue for $H_{\rm rel }$ precisely means that   $(Q {\bf M_{00}}Q)$ is invertible on $ Q L^2(\ere^2) $ with bounded inverse.

Finally, we designate,
\beq \label{2.8b}
Y_0(\nu):= \frac{1}{\sqrt{2\pi}}, \nu \in \ese^1.
\ene

 For $X,Y$ Banach spaces we denote by  ${\mathcal B}\left(X;Y\right)$ the Banach space of all bounded linear operators from  $X$, into $Y$. In the case $X=Y$ we use the notation $\mathcal B(X)$. By  ${\rm Tr A}$ we designate   the trace of the operator $A$.

\begin{theorem} \label{th2.1}
Suppose that  Assumption \ref{ass1.1} is satisfied and that at zero $H_{\mathrm rel}$ has neither a resonance (half-bound state)  nor an eigenvalue.  Then,  in the norm of ${\mathcal B}\left(L^2(\ese^1)\right)$ we have for $|\p/\hbar| \rightarrow 0$ the expansion,
\beq \label{2.8c}
{\mathcal S}(\p^2/2m) = I + i \pi\,  \frac{1}{ \ln|\p/\hbar|} \, \Sigma+  \left( i \pi (\ln 2-\gamma + \frac{1}{a}) -\frac{\pi^2}{2} \right)\,  \frac{1}{|\ln (|\p|/\hbar)|^2} \, \Sigma+
O\left( \frac{1}{|\ln (|\p|/\hbar)|^3}\right),
\ene
where $I$ is the identity operator on $L^2(\ese^1)$,
\beq \label{2.9}
\Sigma:=  \left(\cdot,Y_0 \right) \,Y_0,
\ene
and $a$ is the scattering length defined by
\beq\label{2.10}
\frac{1}{a}:=    \frac{2 \pi}{ \alpha }\,  {\rm Tr}\,\left[ P {\bf N_{00}} P-P {\bf M_{00}} Q D_{00}{\bf M_{00}}+ P {\bf M_{00}}Q \right].
\ene
\end{theorem}
\noindent {\it Proof:}
Let us denote by ${\mathcal S}_1(\lambda)$ the scattering matrix for the Hamiltonian $ H_1:= - \Delta +\frac{2m}{\hbar^2}
V(\x)$. It follows from an elementary argument that,
\beq \label{2.11}
{\mathcal S}(\p^2/2m)= {\mathcal S}_1(\left(\p/\hbar\right)^2).
\ene
Furthermore \cite{ya},
\beq\label{2.12}
{\mathcal S}_1(\lambda)= I- 2\pi i\, \frac{2m}{\hbar^2} \,\Gamma(\lambda) \,v \,\left(M(\lambda)\right)^{-1}\, v\, \Gamma^\ast(\lambda),
\ene
where $\Gamma(\lambda)$ is the trace operator,
\beq \label{2.13}
\left(\Gamma(\lambda) \varphi\right)(\nu):= \frac{1}{\sqrt{2}}\, \frac{1}{2\pi}\, \int_{\ere^2}\, e^{-i \sqrt{\lambda}\,\nu\cdot \x}\,
\varphi(\x)\, d\x, \quad \nu \in \ese^1,
\ene
and
\beq\label{2.14}
M(\lambda):= \left( U+ \frac{2m}{\hbar^2}v (-\Delta-\lambda -i0)^{-1}\, v \right)
\ene
has a bounded inverse in $L^2(\ere^2)$.

Moreover, for all $f \in L^2(\ese^1)$,
\beq
\left\|\, v(\x)\,\left(\Gamma^\ast -\frac{1}{\sqrt{2}}\,\Sigma\right)\, f\right\|_{L^2(\ere^2)} \leq \sqrt{\lambda}\, \frac{1}{2 \sqrt{\pi}}\,\left\| v(\x) \,|\x|\right\|_{L^2(\ere^2)} \,
\|f\|_{L^2(\ese^1)}.
\ene
Furthermore, by Schwarz's inequality,
$$
\left\| v(\x) \,|\x|\right\|^2_{L^2(\ere^2)}= \left\| V(\x) \,|\x|^2\right\|_{L^1(\ere^2)} \leq  \left\|(1+|x|)^{\beta} V(\x) \,\right\|_{L^2(\ere^2)} \left\|(1+|\x|)^{-\beta+2}\right\|_{L^2(\ere^2)} \leq C,
$$
and it follows that,
\beq \label{2.15}
v(\x)\,\Gamma^\ast(\lambda) = \frac{1}{\sqrt{2}}\, v(\x)\, \Sigma + O(\sqrt{\lambda}), \quad \lambda \rightarrow 0,
\ene
in the  operator norm in $ \mathcal B( L^2(\ese^1),L^2(\ere^2))$.
Taking the adjoint in both sides of  \eqref{2.15}    we obtain that,
\beq \label{2.16}
 \Gamma(\lambda)\, v(\x)  = \frac{1}{\sqrt{2}}  \frac{1}{2\pi}\,  \left(\cdot, v\right)+ O(\sqrt{\lambda}), \quad \lambda \rightarrow 0,
\ene
in the operator norm in $ \mathcal B(L^2(\ere^2),  L^2(\ese^1)).$ By \eqref{2.12}, \eqref{2.15}, \eqref{2.16},
\beq\label{2.17}
{\mathcal S}_1(\lambda)= I-  \frac{i}{2}\, \frac{2 m}{\hbar^2} \, \,\left( \left(M(\lambda)\right)^{-1}\, v,\, v \,\right)\, \Sigma\,+ O(\sqrt{\lambda}), \quad  \lambda \rightarrow 0,
\ene
in the norm of ${\mathcal B}\left(L^2(\ese^1)\right)$. Equation \eqref{2.8c} follows from \eqref{2.17} and Theorem 6.2 of
\cite{jn}.

\bull

The low-energy expansion (\ref{2.8c}) was previously proved by \cite{bgd} in the case of exponentially decreasing potentials such that $\int V(\x)\neq 0$.

\section{ The Creation of Entanglement at Low-Energy}\sss
As mentioned in the introduction, we consider  a pure state of the two-particle system. The wave function in the momentum representation is given by  $\varphi(\p_1,\p_2)$.  We designate by $\rho(\varphi)$ the one-particle reduced density matrix with integral kernel,
$$
\rho(\varphi)(\p_1,\p_1'):=\int \varphi(\p_1,\p_2)\,  \overline{ \varphi(\p_1',\p_2)}\,\, d \p_2.
$$

The purity, ${\mathcal P}(\varphi)$, is given by,
\beq\label{3.1}
{\mathcal P}(\varphi):=\hbox{\rm Tr}(\rho^2)= \int \, d\p_1 \,d\p_1' \,d\p_2 \,d\p_2' \,\varphi(\p_1,\p_2) \,\overline{\varphi(\p_1', \p_2)} \,\varphi(\p_1',\p_2') \,\overline{\varphi(\p_1,\p_2')}.
\ene
As is well known \cite{ai, iko, lsa}, the purity is a measure of entanglement   that is  related closely to the  R\'enyi entropy of order $2$,$ - \hbox{\rm ln Tr}( \rho^2)$. Furthemore,  it  has a trivial relation with the linear entropy, $S_L$, given by $S_L=1-\mathcal P$.  Clearly, It satisfies $ 0 \leq \mathcal P \leq 1$ if $\varphi$ is normalized to one. Furthermore, it is equal to one for a product state, $\varphi= \varphi_1(\p_1) \, \varphi_2(\p_2)$. As we will show, the purity can be directly computed in terms of the scattering matrix. For this reason it  is a measure of entanglement that is convenient for the study of entanglement creation in scattering processes.

 As we already said, we consider an incoming asymptotic state,  in the  center-of-mass frame, that is a product of two normalized Gaussian wave functions,
\beq\label{3.2}
\varphi_{\rm in,\p_0}(\p_1,\p_2):=  \varphi_{\p_0}(\p_1) \,  \varphi_{-\p_0}(\p_2),
\ene
where
\beq\label{3.3}
\varphi_{\p_0}(\p_1):=  \frac{1}{(\sigma^2\pi)^{1/2}}   e^{- (\p_1-\p_0)^2 /  2\sigma^2}.
\ene

 Observe that by (\ref{2.1}) the mean value of the relative momentum in the state (\ref{3.2}) is equal to $\p_0$.

Since the incoming asymptotic state $\varphi_{\rm in, \p_0}$ is a product state its purity is one,
\beq\label{3.4b}
\mathcal P(\varphi_{\rm in,\p_0})=1.
\ene
The outgoing  asymptotic state of the two particles, $\varphi_{\rm out, \p_0}$ -after the scattering process is over- is given by
 \beq\label{3.5}
 \varphi_{\rm out,\p_0}(\p_1,\p_2):= \left( \mathcal S( \p^2/2m ) \varphi_{\rm in, \p_0}\right)(\p_1,\p_2).
\ene
As the relative momentum $\p$  depends on $\p_1$ and on $\p_2$, $\varphi_{\rm out,\p_0}$ is not a product state, and then it  has purity smaller than one. This implies that the scattering process has created  entanglement between the two particles.

Let us  introduce some notations that we use later.

Let us designate  by $\varphi_{\rm in}$ the incoming asymptotic state  with mean value of the relative momentum zero,
\beq\label{3.6}
\varphi_{\rm in}(\p_1,\p_2):=  \varphi(\p_1) \,  \varphi(\p_2),
\ene
with,
\beq\label{3.7}
\varphi(\p):=  \frac{1}{(\sigma^2\pi)^{1/2}}   e^{-  \p^2 /  2\sigma^2},
\ene
and by  $\varphi_{\rm out}$  the outgoing asymptotic state with incoming asymptotic state  $\varphi_{\rm in}$,
\beq\label{3.7b}
\varphi_{\rm out}(\p_1,\p_2):=  \left( \mathcal S( \p^2/2m ) \varphi_{\rm in}\right)(\p_1,\p_2).
 \ene

Recall that,
\beq \label{3.7c}
\mu_i= \frac{m_i}{m_1+m_2}, i=1,2,
\ene
is the ratio of the mass of the $i$ particle to the total mass.

It follows from (\ref{2.1}) that,
\beq \label{3.7d}
\p_1= \mu_1 \p_{\rm cm}+\p,
\ene
\beq \label{3.7e}
\p_2= \mu_2 \p_{\rm cm}-\p.
\ene

We prepare  some results that we will use.
It is a consequence of  (\ref{3.2}, \ref{3.3}, \ref{3.7d}, \ref{3.7e}) that
\beq \label{3.11a}
\varphi_{\rm in, \p_0}= \frac{1}{\sigma^2\pi}\, e^{-(\mu_1^2+\mu_2^2) \p_{\rm cm}^2/2\sigma^2}\,\,e^{-(\p-\p_0)^2/\sigma^2}\, e^{-(\mu_1-\mu_2) \p_{\rm cm} \cdot (\p-\p_0)}.
\ene
\begin{remark}
For some positive constant $\delta$,
\beq\label{3.11b}
\left|  \varphi_{\rm in}\right| \leq \,  \frac{1}{\sigma^2\pi}\,  e^{-\delta(\p_{\rm cm}^2+ \p^2)/2 \sigma^2}.
\ene

\noindent {\it Proof:}{ \rm
Note that $ \mu_1^2+\mu_2^2= \frac{1}{2}\left(1+ (\mu_1-\mu_2)^2\right)$. Then, for $\alpha \geq 0$,
\beq \label{3.11c}
\begin{array}{l}
(\mu_1^2+\mu_2^2) \p_{\rm cm}^2 /2 \sigma^2+ \p^2 / \sigma^2+(\mu_1-\mu_2) \p_{\rm cm}\cdot \p  \geq \\\\
\left( \frac{1}{4}+ (\mu_1-\mu_2)^2 (\frac{1}{4}-\frac{\alpha}{2})\right) \frac{\p^2_{\rm cm}}{\sigma^2}+ \frac{\p^2}{\sigma^2} \left(1-\frac{1}{2 \alpha}  \right) \geq \delta(\p_{\rm cm}^2+ \p^2)/2 \sigma^2,
\end{array}
\ene
provided that  we choose $\alpha$ so that, $ 0 < \delta/2 \leq {\rm min} [   \frac{1}{4}+ (\mu_1-\mu_2)^2(\frac{1}{4}-\frac{\alpha}{2}), (1- \frac{1}{2 \alpha})]$.
The remark follows from (\ref{3.11a}) and (\ref{3.11c}).}
\end{remark}

\begin{prop} \label{prop3.1}
For any $ \alpha, \beta  \geq 0$,
\beq \label{3.12}
\left\| \frac{\left(\ln(2+|\p|/\hbar)\right)^\alpha}{(1+|\ln(|\p|/\hbar)|)^{\beta} } \,\,\ds\varphi_{\rm in,\p_0} \right\| =O\left(\frac{1}{|\ln(\sigma/\hbar+|\p_0|/\hbar)|^{\beta}}\right), \qquad \rm{as} \, \sigma/\hbar+|\p_0|/\hbar \rightarrow 0,
\ene

\beq \label{3.12b}
\left\| \frac{\left(\ln(2+|\p|/\hbar)\right)^\alpha}{|\ln(|\p|/\hbar)|^{\beta} } \,\,\ds\varphi_{\rm in} \right\| =O\left(\frac{1}{|\ln(\sigma/\hbar)|^{\beta}}\right), \qquad \rm{as} \, \,\sigma/\hbar \rightarrow 0.
\ene
Furthermore, uniformly for $|\p_0|/\sigma$ in bounded sets,
\beq \label{3.13}
\left\|   \frac{\left(\ln(2+|\p|/\hbar)\right)^\alpha}{(1+|\ln(|\p|/\hbar)|)^{\beta} } \,\,\ds\left( \varphi_{\rm in,\p_0} -\varphi_{\rm in}\right) \right\| \leq \frac{|\p_0|}{\sigma}\, O\left(\frac{1}{|\ln(\sigma/\hbar+|\p_0|/\hbar)|^{\beta}}\right), \qquad \rm{as} \, \sigma/\hbar+|\p_0|/\hbar \rightarrow 0.
\ene
\end{prop}
\noindent {\it Proof:} By (\ref{3.11b}) we have that,

\beq \label{3.13b}
 \left\| \frac{(\ln(2+|\p|/\hbar))^\alpha}{(1+|\ln(|\p|/\hbar)|)^{\beta} } \,\,\ds\varphi_{\rm in,\p_0} \right\|^2 \leq  I_1+I_2,
 \ene
 where, for some $1 > \gamma >0$,
 \beq \label{3.13c}
 I_1:= \int_{|\p|/\sigma \geq  1/  (\sigma/\hbar+|\p_0|/\hbar)^\gamma}\, \left( \frac{(\ln(2+|\p+\p_0|/\hbar))^\alpha}{(1+|\ln(|\p+\p_0|/\hbar)|)^{\beta} }\right)^2
    \frac{1}{\sigma^4\pi^2}\,  e^{-\delta(\p_{\rm cm}^2+ \p^2)/ \sigma^2}\, d\p_{\rm cm}\ d\p \leq C_N (\sigma/\hbar+|\p_0|/\hbar)^N,  \ene
$ N=1,2,\cdots$,  and
\beq \label{3.13d}
 I_2:= \int_{|\p|/\sigma \leq  1/  (\sigma/\hbar+|\p_0|/\hbar)^\gamma}\, \left( \frac{(\ln(2+|\p+\p
 _0|/\hbar))^\alpha}{(1+|\ln(|\p+\p_0|/\hbar)|)^{\beta} }\right)^2
    \frac{1}{\sigma^4\pi^2}\,  e^{-\delta(\p_{\rm cm}^2+ \p^2)/ \sigma^2}\, d\p_{\rm cm}\ d\p.
\ene
Moreover, for  $|\p|/\sigma \leq  1/ (\sigma/\hbar+|\p_0|/\hbar)^\gamma$ and   $(\sigma/\hbar+ |\p_0|/\hbar)^{1-\gamma}  \leq 1/2$,

$$
\frac{1}{1+|\ln(|\p+\p_0|/\hbar)| } \leq    \frac{1}{1+(1-\gamma)  |\ln( \sigma/\hbar+|\p_0|/\sigma) \,\,2^{1/(1-\gamma})| },
$$
and then,
\beq \label{3.13e}
I_2 =O\left( \frac{1}{|\ln(\sigma/\hbar+|\p_0|/\hbar)|^{\beta}}\right), \qquad \sigma/\hbar+|\p_0|/\hbar \rightarrow 0.
\ene
Equation (\ref{3.12}) follows from (\ref{3.13c}) and (\ref{3.13e}).

In the same way, it follows from (\ref{3.11b})  that,

\beq \label{3.13b1}
 \left\| \frac{(\ln(2+|\p|/\hbar))^\alpha}{|\ln(|\p|/\hbar)|^{\beta} } \,\,\ds\varphi_{\rm in} \right\|^2 \leq  I_1+I_2,
 \ene
 where, for some $1 > \gamma >0$,
 \beq \label{3.13c1}
 I_1:= \int_{|\p|/\sigma \geq  1/  (\sigma/\hbar)^\gamma}\, \left( \frac{(\ln(2+|\p|/\hbar))^\alpha}{|\ln(|\p|/\hbar)|^{\beta} }\right)^2
    \frac{1}{\sigma^4\pi^2}\,  e^{-\delta(\p_{\rm cm}^2+ \p^2)/ \sigma^2}\, d\p_{\rm cm}\ d\p \leq C_N (\sigma/\hbar)^N,
     \qquad N=1,2,\cdots,
 \ene
and
\beq \label{3.13d1}
 I_2:= \int_{|\p|/\sigma \leq  1/  (\sigma/\hbar)^\gamma}\, \left( \frac{(\ln(2+|\p|/\hbar))^\alpha}{|\ln(|\p|/\hbar)|^{\beta} }\right)^2
    \frac{1}{\sigma^4\pi^2}\,  e^{-\delta(\p_{\rm cm}^2+ \p^2)/ \sigma^2}\, d\p_{\rm cm}\ d\p.
\ene
As above, for  $|\p|/\sigma \leq  1/ (\sigma/\hbar)^\gamma$ and   $(\sigma/\hbar)^{1-\gamma}  \leq 1$,

$$
\frac{1}{|\ln|\p/\hbar||} \leq    \frac{1}{(1-\gamma)  |\ln( \sigma/\hbar)| },
$$
and then,
\beq \label{3.13e1}
I_2 =O\left( \frac{1}{|\ln(\sigma/\hbar)|^{\beta}}\right), \qquad \sigma/\hbar  \rightarrow 0.
\ene
Equation (\ref{3.12b}) follows from (\ref{3.13c1}) and (\ref{3.13e1}).

We now prove (\ref{3.13}). We first consider the case when $|\p_0|/\sigma \leq 1$.
By (\ref{3.11a}, \ref{3.11b}),

\beq \label{3.14} \begin{array}{l}
\left|\varphi_{\rm in,\p_0} -\varphi_{\rm in}\right|\leq  \frac{1}{\sigma^2\pi}\,  e^{-\delta(\p_{\rm cm}^2+ \p^2)/2 \sigma^2}
\left| e^{-(\p_0^2 +2 \p\cdot \p_0+ (\mu_1-\mu_2) \p_{\rm cm}\cdot \p_0)/ \sigma^2  }-1  \right|\\\\
\leq  \frac{1}{\sigma^2\pi}\,  e^{-\delta(\p_{\rm cm}^2+ \p^2)/2 \sigma^2 } e^{(\p_0^2 +2 |\p|| \p_0|+ |(\mu_1-\mu_2)| |\p_{\rm cm}| |\p_0|)/ \sigma^2  }\,(\p_0^2 +2 |\p|| \p_0|+ |(\mu_1-\mu_2)| |\p_{\rm cm}| |\p_0|)/ \sigma^2 \leq \\\\
 \frac{1}{\sigma^2\pi}\,  e^{-\delta(\p_{\rm cm}^2+ \p^2)/2 \sigma^2 } e^{(1 +2 |\p|/\sigma+  |\p_{\rm cm}|/ \sigma)  }\,
 |\p_0/\sigma|\left( 1+2 |\p|/\sigma+ |\p_{\rm cm}|/\sigma \right).
\end{array}
\ene
Then,
$$
\begin{array}{l}
\left\| \frac{\left(\ln(2+|\p|/\hbar)\right)^\alpha}{(1+|\ln(|\p|/\hbar)|)^{\beta} } \,\,\ds\left( \varphi_{\rm in,\p_0} -\varphi_{\rm in}\right) \right\|^2 \leq  \frac{1}{\sigma^4\pi^2}\,  \int  \frac{\left(\ln(2+|\p|/\hbar)\right)^{2\alpha}}{(1+|\ln(|\p|/\hbar)|)^{2\beta} } \,  e^{-\delta(\p_{\rm cm}^2+ \p^2)/ \sigma^2 }\\\\
 e^{2(1 +2 |\p|/\sigma+  |\p_{\rm cm}|/ \sigma)  }\,
 |\p_0/\sigma|^2 \left( 1+2 |\p|/\sigma+ |\p_{\rm cm}| /\sigma \right)^2 \, d \p_{\rm cm}\, d \p.
 \end{array}
$$

Estimating as in equations (\ref{3.13b}-\ref{3.13e}) with $\p_0=0$, we prove that for $|\p_0|/\sigma \leq 1$,
$$
\left\| \frac{\left(\ln(2+|\p|/\hbar)\right)^\alpha}{(1+|\ln(|\p|/\hbar)|)^{\beta} } \,\,\ds\left( \varphi_{\rm in,\p_0} -\varphi_{\rm in}\right) \right\|
\leq |\p_0/\sigma|\,\, O\left(\frac{1}{|\ln(\sigma/\hbar)|^{\beta}}\right)
\leq |\p_0/\sigma|\,\, O\left(\frac{1}{|\ln(\sigma/\hbar+|\p_0|/\hbar)|^{\beta}}\right),  \rm{as} \,\,\, \sigma/\hbar+|\p_0|/\hbar \rightarrow 0.
$$

    In the case $|\p_0/\sigma| \geq 1$ the estimate is immediate from (\ref{3.12}), because,
    $$
  \begin{array}{l}
\left\| \frac{\left(\ln(2+|\p|/\hbar)\right)^\alpha}{(1+|\ln(|\p|/\hbar)|)^{\beta} } \,\,\ds\left( \varphi_{\rm in,\p_0} -\varphi_{\rm in}\right) \right\|
\leq   \left\|\frac{\left(\ln(2+|\p|/\hbar)\right)^\alpha}{(1+|\ln(|\p|/\hbar)|)^{\beta} } \,\,\ds \varphi_{\rm in,\p_0}\right\|+
\left\|\frac{\left(\ln(2+|\p|/\hbar)\right)^\alpha}{(1+|\ln(|\p|/\hbar)|)^{\beta} } \,\,\ds \varphi_{\rm in}\right\|=
O\left(\frac{1}{|\ln(\sigma/\hbar+|\p_0|/\hbar)|^{\beta}}\right)
\\\\
\leq |\p_0/\sigma| \,\,\,  O\left(\frac{1}{|\ln(\sigma/\hbar+|\p_0|/\hbar)|^{\beta}}\right),
 \qquad \rm{as} \, \,\, \sigma/\hbar+|\p_0|/\hbar \rightarrow 0.
 \end{array}
$$

\bull

We define,
 \beq\label{3.23}
 \mathcal T(\p^2/2m):=\mathcal S(\p^2/m) -I+  i \pi\,  \frac{1}{1+| \ln(|\p|/\hbar)|} ,
 \ene
 where $I$ is the identity operator on $L^2(\ese^1)$.
 It follows from (\ref{2.8c}) and since $ \| \mathcal S(\p^2/2 m)\|_{\mathcal B(L^2(\ese^1))}=1$, that
 \beq\label{3.24}
 \left\| \mathcal T(\p^2/2 m) \right\|_{\mathcal B(L^2(\ese^1))}\leq C  \frac{\left(\ln(2+|\p|/\hbar)\right)^2}{(1+|\ln(|\p|/\hbar)|)^2 } .
 \ene
 Hence by \eqref{3.12},
 \beq\label{3.25}
 \left\| \mathcal T(\p^2/2m)  \varphi_{\rm in, \p_0}\right\|= O\left(\frac{1}{|\ln(\sigma/\hbar+|\p_0|/\hbar)|^{2}}\right), \qquad \rm{as} \, \sigma/\hbar+|\p_0|/\hbar \rightarrow 0.
\ene

 Let us denote,
 \beq \label{3.26}
 \mathcal L(\phi_1,\phi_2,\phi_3,\phi_4):= \int\, \, d\p_1 \,d\p_1' \,d\p_2 \,d\p_2' \,\phi_1(\p_1,\p_2) \,\overline{\phi_2(\p_1', \p_2)} \,\phi_3(\p_1',\p_2') \,\overline{\phi_4(\p_1,\p_2')}.
\ene
We have that,
$$
\mathcal P(\phi)= \mathcal L(\phi,\phi,\phi,\phi).
$$
 The Schwarz inequality implies that,
 \beq \label{3.27}
 \left| {\mathcal L} (\phi_1,\phi_2,\phi_3,\phi_4 ) \right | \leq \Pi_{j=1}^4 \|\phi_j \|.
 \ene
 We state below   our first  result in the low-energy behavior of the purity.
 \begin{theorem}\label{theor3.2}
 Suppose that  Assumption \ref{ass1.1}
 is satisfied and that at zero $H_{\mathrm rel}$ has neither a resonance (half-bound state)  nor an eigenvalue. Then,  uniformly for $|\p_0|/\sigma$ in bounded sets,
 \beq\label{3.28}
 \mathcal P(\varphi_{\rm out, \p_0})= \mathcal P(\varphi_{\rm out})+\frac{|\p_0|}{\sigma}\,\, O\left(\frac{1}{|\ln(\sigma/\hbar+|\p_0|/\hbar)|^{2}}\right), \qquad \rm{as} \, \sigma/\hbar+|\p_0|/\hbar \rightarrow 0.
\ene
 \end{theorem}
 \noindent{\it  Proof:}  Writing $\varphi_{\rm out,\p_0}$ as,
 $$
 \varphi_{\rm out,\p_0}:=  \mathcal S(\p^2/2m) \varphi_{\rm in,\p_0}=  \varphi_{\rm in,\p_0} -i \pi\,  \frac{1}{1+| \ln(|\p|/\hbar)|}
  \varphi_{\rm in,\p_0}+ \mathcal T(\p^2/2m)\varphi_{\rm in,\p_0},
 $$
and using (\ref{3.4b}),  we see that we can write $\mathcal P(\varphi_{\rm out, \p_0})$ as follows,
 \beq\label{3.29}
 \mathcal P(\varphi_{\rm out, \p_0})= 1+ \sum_{i=1}^4 \mathcal L_{1,i}(\p_0, \psi_1,\psi_2,\psi_3,\psi_4)+\mathcal R(\p_0),
 \ene
 where
 \beq\label{3.29b}
\mathcal L_{1,i}(\p_0, \psi_1,\psi_2, \psi_3,\psi_4)= \mathcal L(\psi_1,\psi_2, \psi_3,\psi_4),
\ene
where one of the $\psi_j$ is equal to  $ \mathcal T(\p^2/2m) \varphi_{\rm in,\p_0}$ and the remaining $3$ are equal to $ \varphi_{\rm in, \p_0}$. Similarly,
 \beq\label{3.30}
 \mathcal R(\p_0):= \sum_{i=1}^A\mathcal L_{2,i}(\p_0,\psi_1,\psi_2, \psi_3,\psi_4),
 \ene
for some  integer $A$, and  where each of the $\mathcal L_{2,i}(\p_0, \psi_1,\psi_2, \psi_3,\psi_4)$ is equal to,
\beq\label{3.31}
\mathcal L_{2,i}(\p_0, \psi_1,\psi_2, \psi_3,\psi_4)= \mathcal L(\psi_1,\psi_2, \psi_3,\psi_4),
\ene
where for some $ 2 \leq k \leq 4$,  $k$ of the   $\psi_j$  are equal either to   $- i \pi\,  \frac{1}{1+| \ln|\p/\hbar||}   \varphi_{\rm in, \p_0}$ or to     $ \mathcal T(\p^2/2m) \varphi_{\rm in,\p_0}$ and the remaining $4-k$ are equal to $ \varphi_{\rm in, \p_0}$. Similarly,
\beq\label{3.32}
 \mathcal P(\varphi_{\rm out })= 1+ \sum_{i=1}^4 \mathcal L_{1,i}(0, \psi_1,\psi_2,\psi_3,\psi_4) + \mathcal R(0),
 \ene
 with
\beq\label{3.33}
 \mathcal R(0):= \sum_{i=1}^A\mathcal L_i(0,\psi_1,\psi_2, \psi_3,\psi_4).
 \ene
Below we prove that,
\beq\label{3.33b}
\mathcal  L_{1,i}(\p_0, \psi_1,\psi_2, \psi_3,\psi_4)= \mathcal  L_{1,i}(0, \psi_1,\psi_2, \psi_3,\psi_4)+ \frac{|\p_0|}{\sigma}\,  O\left(\frac{1}{|\ln(\sigma/\hbar+|\p_0|/\hbar)|^{2}}\right), \quad i=1,2,3,4,
\ene
\beq\label{3.34}
  \mathcal R(\p_0)= \mathcal R(0)+\frac{|\p_0|}{\sigma}\,\,  O\left(\frac{1}{|\ln(\sigma/\hbar+|\p_0|/\hbar)|^{2}}\right),
  \ene
  what proves the theorem in view of (\ref{3.29},\ref{3.32}).

  We proceed to prove (\ref{3.33b}). Without losing generality we can assume that,
  \beq\label{3.35}
  \mathcal L_{1,1}(\p_0,\psi_1,\psi_2, \psi_3,\psi_4)= \mathcal L(\mathcal T(\p^2/2m) \varphi_{\rm in,\p_0},\varphi_{\rm in, \p_0},\varphi_{\rm in, \p_0},\varphi_{\rm in, \p_0}).
 \ene
  We have that,
   \beq\label{3.36}
   \begin{array}{l}
  \mathcal L_{1,1}(\p_0,\psi_1,\psi_2, \psi_3,\psi_4)= \mathcal L(\mathcal T(\p^2/2m) \varphi_{\rm in},\varphi_{\rm in, \p_0},\varphi_{\rm in, \p_0},\varphi_{\rm in, \p_0})+\\\\
  \mathcal L(\mathcal T(\p^2/2m) (\varphi_{\rm in,\p_0}-\varphi_{\rm in}),\varphi_{\rm in, \p_0},\varphi_{\rm in, \p_0},\varphi_{\rm in, \p_0}).
  \end{array}
 \ene
 By (\ref{3.13}, \ref{3.24}, \ref{3.27}, \ref{3.36}),
 \beq\label{3.37}
 \begin{array}{l}
  \mathcal L_{1,1}(\p_0,\psi_1,\psi_2, \psi_3,\psi_4)=   \mathcal L(\mathcal T(\p^2/2m) \varphi_{\rm in},\varphi_{\rm in, \p_0},\varphi_{\rm in, \p_0},\varphi_{\rm in, \p_0})+ \frac{|\p_0|}{\sigma}\, O\left(\frac{1}{|\ln(\sigma/\hbar+|\p_0|/\hbar)|^{2}}\right).
  \end{array}
  \ene
  In the same way, using (\ref{3.13}, \ref{3.25}, \ref{3.37}), we prove that,
  \beq\label{3.38}
  \mathcal L_{1,1}(\p_0,\psi_1,\psi_2, \psi_3,\psi_4)=   \mathcal L(\mathcal T(\p^2/2m) \varphi_{\rm in},\varphi_{\rm in},\varphi_{\rm in, \p_0},\varphi_{\rm in, \p_0})+  \frac{|\p_0|}{\sigma}\,  O\left(\frac{1}{|\ln(\sigma/\hbar+|\p_0|/\hbar)|^{2}}\right).
  \ene
  Repeating this argument two more times we obtain that,
    \beq
  \mathcal L_{1,1}(\p_0,\psi_1,\psi_2, \psi_3,\psi_4)=
  \mathcal L_{1,1}(0,\psi_1,\psi_2, \psi_3,\psi_4)+  \frac{|\p_0|}{\sigma}\,  O\left(\frac{1}{|\ln(\sigma/\hbar+|\p_0|/\hbar)|^{2}}\right).
  \ene

We prove in the same way that (\ref{3.33b}) holds for $i=2,3,4$.   Furthermore, (\ref{3.34}) is proven by the same argument.

  \bull

The next theorem gives us  the leading order of the purity of  $\varphi_{\rm out }$ at low-energy.

 \begin{theorem}\label{theor3.3}
 Suppose that  Assumption \ref{ass1.1} is satisfied and that at zero $H_{\mathrm rel}$ has neither a resonance (half-bound state) nor an eigenvalue. Then, as $ \sigma/\hbar  \rightarrow 0$.

 \beq\label{3.43}
 \mathcal P(\varphi_{\rm out})= \mathcal P\left(\left[  I + i \pi\,  \frac{1}{ \ln|\p/\hbar|} \, \Sigma+  \left( i \pi (\ln 2-\gamma + \frac{1}{a}) -\frac{\pi^2}{2} \right)\,  \frac{1}{(\ln |\p/\hbar|)^2} \, \Sigma\right]\varphi_{\rm in }\right)+
  O\left(\frac{1}{|\ln(\sigma/\hbar)|^{3}}\right).  \ene
 \end{theorem}
 \noindent{\it  Proof:} We write  $\varphi_{\rm out}$ as follows,
 $$
 \varphi_{\rm out}=  \varphi_{\rm out, 1}+ \mathcal T_1(\p^2/2m)  \varphi_{\rm in},
$$
where,
\beq\label{3.44}
\varphi_{\rm out, 1}:= \left[ I + i \pi\,  \frac{1}{ \ln|\p/\hbar|} \, \Sigma+  \left( i \pi (\ln 2-\gamma + \frac{1}{a}) -\frac{\pi^2}{2} \right)\,  \frac{1}{(\ln |\p/\hbar|)^2} \, \Sigma \right]\varphi_{\rm in },
\ene
and
$$
{\mathcal T}_1:= \mathcal S(\p^2/2m) - I - i \pi\,  \frac{1}{ \ln|\p/\hbar|} \, \Sigma-  \left( i \pi (\ln 2-\gamma + \frac{1}{a}) -\frac{\pi^2}{2} \right)\,  \frac{1}{(\ln |\p/\hbar|)^2} \, \Sigma.
$$
By  (\ref{2.8c}) and since $ \| \mathcal S(\p^2/2 m)\|_{\mathcal B(L^2(\ese^1))}=1$,
\beq\label{3.44b}
 \left\| {\mathcal T}_1(\p^2/2 m) \right\|_{\mathcal B(L^2(\ese^1))}\leq C  \frac{\left(\ln(2+|\p|/\hbar)\right)^3}{|\ln(|\p|/\hbar)|^3 } .
 \ene

Using this decomposition we  write $\mathcal P(\varphi_{\rm out})$ as follows,
 \beq\label{3.45}
 \mathcal P(\varphi_{\rm out})= \mathcal P(\varphi_{\rm out, 1})+ \mathcal R_1(\sigma),
 \ene
 where $\mathcal R_1(\sigma)$  is given by,
 \beq\label{3.46}
 \mathcal R_1(\sigma):= \sum_{i=1}^D\mathcal L_i(\sigma,\psi_1,\psi_2, \psi_3,\psi_4),
 \ene
for some  integer $D$, and  where each of the $\mathcal L_i(\sigma, \psi_1,\psi_2, \psi_3,\psi_4)$ is equal to,
\beq\label{3.47}
\mathcal L_i(\sigma, \psi_1,\psi_2, \psi_3,\psi_4)= \mathcal L(\psi_1,\psi_2, \psi_3,\psi_4),
\ene
where for some $ 1 \leq k \leq 4$,  $k$ of the   $\psi_j$  are equal to  $\varphi_{\rm out,1}$ and the remaining $4-k$ are equal to $  \mathcal T_1(\p^2/2m)  \varphi_{\rm in}$.

We complete the proof of the theorem proving that,

\beq\label{3.48}
  \mathcal R_1(\sigma)= O\left(\frac{1}{|\ln(\sigma/\hbar)|^{3}}\right), \qquad \rm{as} \,\,\, \sigma/\hbar  \rightarrow 0.
\ene

  We can assume that,
  \beq\label{3.49}
  \mathcal L_1(\sigma,\psi_1,\psi_2, \psi_3,\psi_4)= \mathcal L(  \varphi_{\rm out, 1},  \varphi_{\rm out, 1}, \varphi_{\rm out, 1}, \mathcal T_1(\p^2/2m) \varphi_{\rm in}).
 \ene
 By (\ref{3.12b}, \ref{3.44b})  we have that,
   \beq\label{3.50}
  \mathcal L_1(\sigma,\psi_1,\psi_2, \psi_3,\psi_4)=O\left(\frac{1}{|\ln(\sigma/\hbar)|^{3}}\right), \qquad \rm{as} \, \sigma/\hbar  \rightarrow 0.
 \ene

We estimate the remaining terms in (\ref{3.48}) in the same way.

  \bull

Let us denote,
$$
\psi(\q):=  \frac{1}{(\pi)^{1/2}}   e^{- \q^2 /  2},  \q \in \ere^2,
$$
$$
\psi_{\rm in}(\q_1,\q_2):=  \psi(\q_1) \,  \psi(\q_2).
$$

\begin{prop} \label{prop3.4}
For any $ \alpha, \beta  \geq 0$,

\beq \label{3.50b}
\left\| \frac{(\ln|\q|)^\alpha}{|\ln(\sigma |\q|/\hbar)|^{\beta} } \,\,\ds\psi_{\rm in} \right\| =O\left(\frac{1}{|\ln(\sigma/\hbar)|^{\beta}}\right), \qquad \rm{as} \, \,\sigma/\hbar \rightarrow 0.
\ene

\end{prop}
\noindent {\it Proof:}  We follow the proof of (\ref{3.12b}). By (\ref{3.11b})  with $\sigma=1$,

\beq \label{3.50c}
 \left\| \frac{(\ln|\q|)^\alpha}{|\ln(\sigma|\q|/\hbar)|^{\beta} } \,\,\ds\psi_{\rm in} \right\|^2 \leq  I_1+I_2,
 \ene
 where, for some $1 > \gamma >0$,
 \beq \label{3.50d}
 I_1:= \int_{|\q| \geq  1/  (\sigma/\hbar)^\gamma}\, \left( \frac{|\ln|\q||^{2\alpha}}{|\ln(\sigma |\q|/\hbar)|^{2\beta} }\right)
    \frac{1}{\pi^2}\,  e^{-\delta(\q_{\rm cm}^2+ \q^2)/ }\, d\q_{\rm cm} \, d\q \leq C_N (\sigma/\hbar)^N,
     \qquad N=1,2,\cdots,
 \ene
and
\beq \label{3.50e}
 I_2:= \int_{|\q| \leq  1/  (\sigma/\hbar)^\gamma}\, \left( \frac{(\ln|\q|)^\alpha}{|\ln(\sigma|\q|/\hbar)|^{\beta} }\right)^2
    \frac{1}{\pi^2}\,  e^{-\delta(\q_{\rm cm}^2+ \q^2)}\, d\q_{\rm cm}\ d\q.
\ene
Furthermore, for  $|\q| \leq  1/ (\sigma/\hbar)^\gamma$ and   $(\sigma/\hbar)^{1-\gamma}  \leq 1$,

$$
\frac{1}{|\ln(\sigma|\q|/\hbar)|} \leq    \frac{1}{(1-\gamma)  |\ln( \sigma/\hbar)| },
$$
and then,
\beq \label{3.50f}
I_2 =O\left( \frac{1}{|\ln(\sigma/\hbar)|^{\beta}}\right), \qquad \sigma/\hbar \rightarrow 0.
\ene
Equation (\ref{3.50b}) follows from (\ref{3.50d}) and (\ref{3.50f}).

\bull

A straightforward computation with the help of (\ref{3.50b}) shows that,
\beq\label{3.55}\begin{array}{l}
\mathcal P\left( \left[  I + i \pi\,  \frac{1}{ \ln|\p/\hbar|} \, \Sigma+  \left( i \pi (\ln 2-\gamma + \frac{1}{a}) -\frac{\pi^2}{2} \right)\,  \frac{1}{(\ln |\p/\hbar|)^2} \, \Sigma\right]\varphi_{\rm in }\right)= 1- \frac{1}{(\ln(\sigma/ \hbar))^2}  \left( \mathcal  P_1(\psi_{\rm in})+\mathcal P_2(\psi_{\rm in})\right)+\\\\
 O\left(\frac{1}{(\ln(\sigma/\hbar))^3}\right), \quad \hbox{\rm as}\, \sigma/\hbar \rightarrow 0,
 \end{array}
\ene
where,
\beq\label{3.56}
\mathcal P_1(\psi_{\rm in})= \Sigma_{j=1}^3 \mathcal P_{1,j}(\psi_{\rm in}),
\ene
with
\beq\label{3.57}
\mathcal P_{1,1}(\psi_{\rm in}) = -2 \pi^2 \int d\q_1d\q_2d\q_3\,
(\Sigma\psi_{\rm in}(\q_1,\q_2))\, (\Sigma\psi_{\rm in}(\q_3,\q_2))\,\psi_{\rm in}(\q_1,\q_3),
\ene
\beq\label{3.58}
\mathcal P_{1,2}(\psi_{\rm in})= -2 \pi^2 \int d\q_1d\q_2d\q_3\,
\left(\Sigma\psi_{\rm in}(\q_1,\q_2)\right)\, \left(\Sigma \psi_{\rm in}(\q_1,\q_3)\right) \psi_{\rm in}(\q_2,\q_3),
\ene
\beq\label{3.59}
\mathcal P_{1,3}(\psi_{\rm in})=2 \pi^2\left[\int d\q_1d\q_2 \,
\left(\Sigma\psi_{\rm in}(\q_1,\q_2)\right)\,  \psi_{\rm in}(\q_1,\q_2)\right]^2,
\ene
and
\beq\label{3.60}
\mathcal P_{2}(\psi_{\rm in})=
2 \pi^2 \int d\q_1d\q_2 \,
\left(\Sigma \psi_{\rm in}(\q_1,\q_2)\right)\, \psi_{\rm in}(\q_1,\q_2).
\ene
Explicitly evaluating the integrals in (\ref{3.57}, \ref{3.58}, \ref{3.59})   using (\ref{3.11a}),  we prove that,
\beq\label{3.61}
\mathcal P_{1,1}(\psi_{\rm in}) = -\frac{2}{\pi}\, J(\mu_1,\mu_2),
\ene
\beq\label{3.62}
\mathcal P_{1,2}(\psi_{\rm in}) = -\frac{2}{\pi}\, J(\mu_2,\mu_1),
\ene
\beq\label{3.63}
\mathcal P_{1,3}(\psi_{\rm in}) = 2 \pi^2\,\,\left( L(\mu_1,\mu_2)\right)^2,
\ene
 \beq\label{3.64}
 \mathcal P_{2}(\psi_{\rm in})= 2 \pi^2 L(\mu_1,\mu_2),
\ene
where,
\beq \label{3.65}\begin{array}{c}
J(\mu_1,\mu_2):=  \int d\q_2\left[ \,\,\int d\q_1\, \hbox{\rm Exp}[-\frac{1}{2}(\mu_1^2+\mu_2^2)(\q_1+\q_2)^2-(\mu_2\q_1-\mu_1\q_2)^2 - \q_1^2/2] \,\, \cdot\right.\\\\\left.
\ds I_0(|\mu_1-\mu_2|\, |\q_1+\q_2|\,|\mu_2\q_1-\mu_1\q_2|)\right]^2.
 \end{array}
\ene
Here $I_0$ is the modified  Bessel function \cite{gr}, and
 \beq \label{3.65b}
L(\mu_1,\mu_2):= \int_0^\infty \, \int_0^\infty\, d\lambda \, d \rho\, e^{-2\lambda}\, e^{- (\mu_1^2+\mu_2^2)\rho}\, \left(
 I_0(|\mu_1-\mu_2| \sqrt{\lambda\, \rho}\,)\right) ^2.
\ene

We prove in the appendix that,
\beq\label{3.66}
L(\mu_1,1-\mu_1)= \frac{1}{\sqrt{1+(2\mu_1-1)^2}}.
\ene

We denote by $\mathcal E(\mu_1$) the entanglement coefficient,
\beq\label{3.67}\begin{array}{l}
\mathcal E(\mu_1):= 2 \pi^2 L(\mu_1,1-mu_1) \left(1+L(\mu_1,1-mu_1)\right) - \frac{2}{\pi} [J(\mu_1,1-\mu_1)+ J(1-\mu_1,\mu_1)]=\\\\ \ds
\frac{2 \pi^2}{1+(2\mu_1-1)^2}\,\, \left[1+ \sqrt{1+(2\mu_1-1)^2}\right]- \frac{2}{\pi} [J(\mu_1,1-\mu_1)+ J(1-\mu_1,\mu_1)].
\end{array}
\ene

The next  theorem is our main result.
\begin{theorem}\label{theor3.6}
 Suppose that  Assumption \ref{ass1.1} is satisfied and that at zero $H_{\mathrm rel}$ has neither a resonance (half-bound state)  nor an eigenvalue.  Then, \beq\label{3.68}
 \mathcal P(\varphi_{\rm out})=1-  \frac{1}{(\ln(\sigma/\hbar))^2} \mathcal E(\mu_1) + O\left(\frac{1}{|\ln(\sigma/\hbar)|^3}\right),  \qquad   \rm{as}\, \sigma/\hbar \rightarrow 0,\ene
 where the entanglement coefficient $\mathcal E(\mu_1)$ is given by (\ref{3.67}).
 \end{theorem}

 \noindent {\it Proof:} The theorem follows from (\ref{3.43}, \ref{3.55}, \ref{3.56}, \ref{3.61}-\ref{3.64}, \ref{3.66}).

 \bull

Note  that $\mathcal E(\mu_1)=\mathcal E(1-\mu_1)$, as it should be, because $\mathcal P(\varphi_{\rm out})$ is invariant under the exchange of particles one and two.

 Observe that,
 $$
 J(1/2,1/2)= \pi^3.
$$
By (\ref{3.67}) for $\mu_1=1/2$, when the masses are equal, the entanglement coefficient is zero, $\mathcal E(1/2)= 0$.
Of course, this only means that in this case the purity is one at  leading order.

We  explicitly evaluate in the appendix  $J(1,0)$,
\beq \label{3.77}
J(1,0)= 16.6377.
\ene
For  $\mu_1 \in [0,1]\setminus\{1/2,1\}$ we compute  $J(\mu_1,1-\mu_1)$ numerically using Gaussian quadratures.
In Table 1 and in Figure 1 we give  values of $\mathcal E(\mu_1)$ for $0.5 \leq \mu_1:= m_1/(m_1+m_2) \leq 1$.

\section{Conclusions} \sss
 In this paper we give a rigorous computation, with error bound, of the entanglement created in the low-energy scattering of two particles in
  two dimensions. The  interaction between the particles is given by   potentials that are not required to be spherically symmetric.
   Before the scattering  the particles are in a pure state that is a product of two  normalized Gaussians with the same variance  $\sigma$.
   After the collision the particles are in a outgoing asymptotic state that is not a product state. The measure of the entanglement created
   by the collision is the purity, $\mathcal P$,  in the state  after the collision. Before the collision the purity
   is one.

 We prove that   $\mathcal P= 1-\, \frac{1}{(\ln (\sigma/\hbar))^2} \, \mathcal E+ O\left(\frac{1}{|\ln(\sigma/\hbar)|^3}\right)$,
  as  $\sigma/\hbar \rightarrow 0$,  where $\sigma$ is the variance of the Gaussians and the entanglement coefficient, $\mathcal E$, depends only
  on the masses of the particles and not on the interaction potential. This proves that the entanglement created at low-energy in two
  dimensions is universal, in the sense that it is independent on the interaction potential between the particles.
  This is strikingly different with the three dimensional case, that we considered in \cite{we}, where  the entanglement created at
   low-energy  is proportional to the total scattering  cross section. However, the entanglement  depends strongly in the difference of
   the masses. As in three dimensions \cite{we} the minimum is taken  when the masses are equal, and  it rapidly increases
   with the difference of the masses.

\section{Appendix} \sss
By \eqref{3.65b} we have that \cite{gr}
\beq \label{a.1}\begin{array}{l}
L(\mu_1, 1-\mu_1)= \int_0^\infty
\, d \lambda \,e^{-2\lambda}\, \frac{2}{1+(2\mu_1-1)^2}\, I_0\left(\frac{(2\mu_1-1)^2 \lambda}{ 1+(2\mu_1-1)^2 }\right) \,
\rm{Exp}\left[ \frac{(2\mu_1-1)^2 \lambda} {1+(2\mu_1-1)^2 }\right]= \ds
\frac{1}{\sqrt{1+(2\mu_1-1)^2} }.
\end{array}
\ene
Moreover, by (\ref{3.65}) and denoting  $\q_{\rm cm}:= \q_1+\q_2$,
\beq\label{a.2}
J(1,0)=\int\, d \q_2\, e^{-3 \q_2^2}\, \left[ \int \,d \q_{\rm cm}\, {\rm Exp}\left(-\q_{\rm cm}^2+ \q_{\rm cm}\cdot \q_2\right)\, I_0(|\q_{\rm cm}|
 |q_2| )\right]^2.
 \ene
Furthermore, using polar coordinates, \cite{gr},
\beq\label{a.3}\begin{array}{l}
J(1,0)=\pi^3\, \int_0^\infty\, d \lambda \,e^{-3\lambda}\, \left[ \int_0^\infty\, d \rho\, e^{-\rho}\, \left( I_0(\sqrt{\rho}\,\sqrt{\lambda})\right)^2\right]^2=\\\\
\pi^3\, \int_0^\infty\, d \lambda \,e^{-2\lambda}\,  \left( I_0(\lambda/2)\right)^2= \pi^2 K(0.25)= 16.6377,
\end{array}
\ene
where $K(x)$ is the complete elliptic integral.

\noindent {\bf Acknowledgement}

\noindent This research was partially done while I was visiting the Departamento de F\'{\i}sica, Facultad de Ciencias Exactas Ingenier\'{\i}a y Agrimensura, Universidad Nacional de Rosario and Instituto de F\'{\i}sica Rosario, Consejo Nacional de Investigaciones Cient\'{\i}ficas y T\'ecnicas, Argentina. I thank Mario Castagnino and Luis Lara for their kind hospitality. I thank Gerardo Dar\'{\i}o Flores Luis  for his help in the numerical computation of $J(\mu_1,1-\mu_1)$.

\newpage

\begin{table}\label{tab1}
 \caption{The Entanglement Coefficient $\mathcal E(\mu_1)$ }
  \begin{center}
  \begin{tabular}{ l c  }
 $ \mu_1:= m_1/(m_1+m_2)$ & $\mathcal E (\mu_1)$  \\
  0.5 &  0.000        \\
  0.525 &   0.0001              \\
  0.55    &   0.0012                      \\
  0.575    &      0.0057                  \\
   0.6   &       0.0174                   \\
   0.625   &     0.0408                \\
   0.65  &      0.0806               \\
    0.675  &        0.1410         \\
    0.7  &             0.2253                 \\
    0.725  &         0.3357                     \\
    0.75  &           0.4725                \\
    0.775  &        0.6348                 \\
    0.8  &             0.8203                \\
     0.825 &          1.0255            \\
     0.85 &               1.2462                 \\
     0.875 &           1.4776       \\
     0.9 &            1.7151              \\
     0.925 &          1.9542          \\
      0.95&         2.1909       \\
      0.975&            2.4216    \\
      1&               2.6436
\end{tabular}
\end{center}
\end{table}
\begin{figure}\label{fig1}
\centering
\setlength{\unitlength}{1cm}
\hspace{30cm}
\includegraphics[width=25cm,totalheight=40cm]{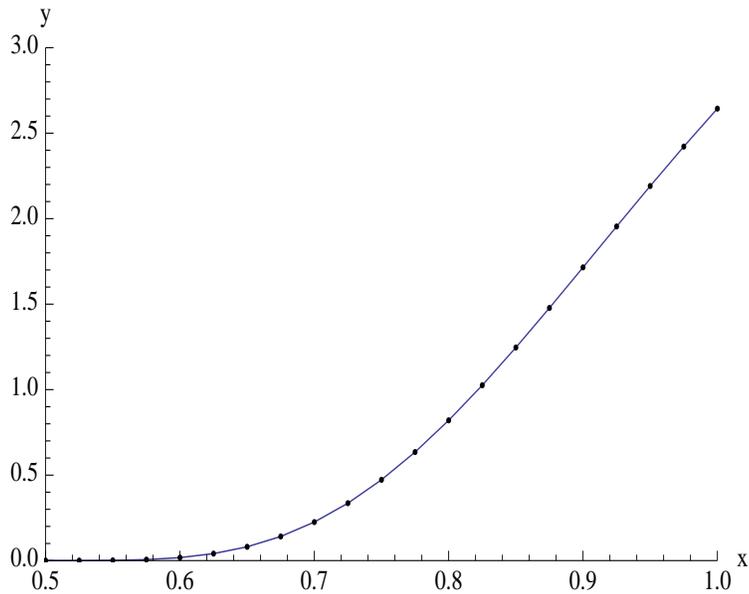}
\vspace{-29cm}
\caption{The entanglement coefficient  $y=\mathcal E(\mu_1)$, as a function of $x=\mu_1= m_1/(m_1+m_2)$,  for $0.5\leq \mu_1 \leq 1$.}
\end{figure}
\end{document}